\documentclass[11pt,a4paper,twocolumn]{article}
\usepackage{amsmath,amssymb}
\usepackage{graphics} 
\usepackage{epsfig}
\newcommand{\comment}[1]{{\relax}}
\begin{document}
\DeclareGraphicsExtensions{.eps} 
\twocolumn[
\begin{center}
{\LARGE \bf Scalar fields superdense gravitating systems}\\
\bigskip
{A. V. Minkevich${}^{1,2}$, A. S. Garkun${}^1$, Yu. G.
Vasilevski${}^1$}\\ {\it ${}^1$Department of Theoretical Physics,
Belarussian State University,\\
 av. F. Skoriny 4, 220050, Minsk, Belarus;\\
 {\small \rm  E-mail: MinkAV@bsu.by, garkun@bsu.by}\\
 ${}^2$Department of
Physics and Computer Methods,\\
 Warmia and Mazury University in
Olsztyn, Poland\\ {\small \rm  E-mail: awm@matman.uwm.edu.pl} }
\bigskip
\begin{minipage}{0.8\textwidth}
{{\bf Abstract}. Solutions for scalar fields superdense
gravitating systems of flat, open and closed type obtained in the
frame of gauge theories of gravitation are discussed. Properties
of these systems in dependence on parameter $\beta$ and initial
conditions are analyzed.}
 {{\large PACS:} 0420J, 0450, 1115, 9880}
 \end{minipage}
\end{center}]

 \section{Introduction}

As it was shown at first in \cite{l1}, generalized cosmological
Friedmann equations (GCFE)  deduced in the frame of gauge theories
of gravity (GTG) \cite{l2} besides cosmological solutions lead to
solutions for some hypothetical objects --- so-called superdense
gravitating systems (SGS). SGS have extremely high energy density
and their dynamics is essentially noneinsteinian. In considered
approximation of homogeneous isotropic space SGS oscillate between
minimum and maximum values of the scale factor. Later solutions
for such systems in the case of various equations of state for
gravitating matter were discussed in [3--5]. Below SGS are
investigated for systems including scalar fields.

\section{Generalized cosmological Friedmann equations}

GCFE for systems including scalar field $\phi$ minimally coupled
with gravitation and radiation have the following form
\cite{l7}\footnote{One supposes the interaction between scalar
field and radiation is negligibly small.}.
\begin{eqnarray}
\lefteqn{\frac{k}{R^2}
Z^2+\left\{H\left[1-2\beta(2V+\dot{\phi}^2)\right]-3\beta
V'\dot{\phi}\right\}^2} \\ \lefteqn{\phantom{H} =\frac{8
\pi}{3M_p^2}\,\left[\rho_r+
\frac{1}{2}\dot{\phi}^2+V-\frac{1}{4}\beta\left(4V-\dot{\phi}^2
\right)^2\right]Z,}\nonumber\\
\lefteqn{\dot{H}\left[1-2\beta(2V+\dot{\phi}^2)\right]Z
}\nonumber\\ & & \phantom{H}+H^2\left\{\left[
1-4\beta(V-4\dot{\phi}^2)\right]Z-18\beta^2\dot{\phi}^4\right\}\nonumber\\
& &\phantom{H} +12\beta
H\dot{\phi}V'\left[1-2\beta(2V+\dot{\phi}^2)\right] \\ & &
\phantom{H}-3\beta\left[(V''\dot{\phi}^2-V'{}^2)Z+6\beta\dot{\phi}^2
V'{}^2\right] \nonumber\\ & &\phantom{H} =\frac{8
\pi}{3M_p^2}\,\left[V-\dot{\phi}^2-\rho_r
-\frac{1}{4}\beta(4V-\dot{\phi}^2)^2\right]Z\nonumber,
\end{eqnarray}
where $R$ is the scale factor, $k=+1,0,-1$ for closed, flat and
open models respectively, $H$ is the Hubble parameter, $\rho_r$ is
radiation energy density, $V(\phi)$ is a scalar field potential,
$\displaystyle V'=\frac{dV}{d\phi}$, $\displaystyle
V''=\frac{d^2V}{d\phi^2}$, $M_p$ is Planckain mass,
$Z=1-\beta(4V-\dot{\phi}^2)$. (The system of units with
$\hbar=c=1$ is used.) Eqs. (1)--(2) include indefinite parameter
$\beta$ with inverse dimension of energy density, the value of
$|\beta|^{-1}$ determines the scale of extremely high energy
densities.

Besides Eqs. (1)--(2) gravitational equations of GTG lead to the
following relation for torsion function $S$ and nonmetricity
function $Q$ \cite{l2,l6}
\begin{equation}
S-\frac{1}{4}Q=\frac{3\beta}{2}\,\frac{\left(H\dot{\phi}+V'\right)\dot{\phi}}{1-\beta
\left(4V-\dot{\phi}^2\right)}.
\end{equation}
The conservation law in usual form  follows from Eqs. (1)--(2), as
result we obtain the equation for scalar field
\begin{equation}
\ddot{\phi}+3H\dot{\phi}=-V'
\end{equation}
and the integral for radiation ${\rho_rR^4={\rm const}}$. If the
condition
\begin{equation}
|\beta(4V-\dot{\phi}^2)|\ll 1
\end{equation}
is valid, solutions of GCFE coincide practically with that of
general relativity (GR) and the torsion and nonmetricity functions
are negligibly small.

As it was shown in \cite{l7}, the GCFE allow to build regular
inflationary cosmological models with dominating ultrarelativistic
matter at a bounce, if the scale of extremely high energy
densities is much less than the Planckian one $|\beta|^{-1}\ll
1M_p^{4}$ ($\beta<0$). Numerical solutions obtained in
\cite{l7,l8} for closed and flat models show that after inflation
($|\phi|<1M_p$) during certain time intervals the dynamics of
inflationary models has oscillating character, namely scalar field
and the Hubble parameter oscillate near values zero. Such
situation is typical for SGS. This means, if $|\phi|<1 M_p$ and
the relation (5) is not fulfilled, by choosing some initial
conditions we can obtain numerical solutions of GCFE for SGS
including scalar fields. We have to take into account that by
given initial conditions for magnitudes ($\phi$, $\dot\phi$,$R$,
$\rho_r$) there are two different solutions corresponding to two
values of the Hubble parameter, which according to (1) are
\[ 
  H_{\pm}=\frac{3\beta V'\dot{\phi}\pm\sqrt{D}}{1-2 \beta(2V+\dot{\phi}^2)},
\] 
where
\begin{eqnarray}
& &\hspace{-2em}D=\frac{8\pi}{3M_p^2}\,\left[\rho_r+
\frac{1}{2}\dot{\phi}^2+V-\frac{1}{4}\beta\left(4V-\dot{\phi}^2
\right)^2\right]Z\nonumber\\ & &-\frac{k}{R^2}\,Z^2\ge 0.\nonumber
\end{eqnarray}
Solutions of GCFE corresponding to $H_{+}$ and $H_{-}$ were called
in \cite{l8} as $H_{+}$-solutions and $H_{-}$-solutions
respectively.

Initial conditions will be chosen at the moment $t=0$
corresponding to extremum of the scale factor $R$: $R(0)=R_0$ and
$H_{+}(0)=0$ or $H_{-}(0)=0$. Then according to (1) initial values
of $R_0$, $\rho_{r0}$, $\phi_0$ and $\dot{\phi}_0$ satisfy the
following relation
\begin{eqnarray}
& &\hspace{-2em}\frac{k}{R_0^2} Z_0^2+9\beta^2
V'{}_0^2\dot{\phi}_0^2\nonumber\\
 & &\hspace{-2em}\phantom{=}=\frac{8
\pi}{3M_p^2}\,\left[\rho_{r0}+\frac{1}{2}\dot{\phi_0}^2
+V_0-\frac{1}{4}\beta\left(4V_0-\dot{\phi}_0^2\right)^2\right]
Z_0,\nonumber\\
\end{eqnarray}
Eq. (6) determines on the plane $P$ with the axes ($\phi$,
$\dot{\phi}$) curves ($H_0$-curves), in points of which the Hubble
parameter vanishes for $H_{+}$-solutions or $H_{-}$-solutions.
From (2) the time derivative of the Hubble parameter $\dot{H}_0$
at the moment $t=0$ is
\begin{eqnarray}
\lefteqn{\dot{H}_0=\left\{\frac{8
\pi}{3M_p^2}\,\left[V_0-\dot{\phi}_0^2-\rho_{r0}-\frac{1}{4}\beta(4V_0
-\dot{\phi}_0^2)^2\right]\right. }\nonumber\\ &
&\left.\phantom{H_0=}+3\beta\left[(V''{}_0\dot{\phi}_0^2-V'{}_0^2)
+6\beta\dot{\phi}_0^2 V'{}_0^2Z_0^{-1}\right]\right\}\nonumber\\ &
& \times\left[1-2\beta(2V_0+\dot{\phi}_0^2)\right]^{-1},
\end{eqnarray}
where $Z_0=1-\beta(4V_0-\dot{\phi}_0^2)$.

In order to analyze solutions properties near the origin of
coordinates on the plane $P$ we will examine $H_0$-curves defined
by (6) for some simple scalar field potentials.

\section{Extremum $H_0$-curves for simplest scalar field
potentials}

According to (1) in the case of flat and closed models we have
$Z\ge 0$ ($\beta<0$) and admissible values of scalar field $\phi$
and derivative $\dot\phi$ on the plane $P$ are limited by two
bounds $L_{\pm}$ defined by equation
\begin{equation}
\dot{\phi}=\pm \left(4V+|\beta|^{-1}\right)^{\frac{1}{2}}.
\end{equation}
$H_0$-curves $B_1$ and $B_2$ for flat models ($k=0$) are situated
on the plane $P$ between bounds $L_{+}$ and $L_{-}$ having with
them common points (points $K_1$ and $K_2$ in Fig.~1) on the axis
$\dot\phi$ (if $V'(0)=0$). Each of $H_0$-curves contains two parts
--- extremum curve for $H_{+}$-solutions ($H_{0+}$-curves) and
extremum curve for $H_{-}$-solutions ($H_{0-}$-curves) \cite{l8}.
The family of $H_0$-curves for closed models is situated on the
plane $P$ between curves $B_1$ and $B_2$. The family of
$H_0$-curves for open models is situated on the plane $P$ between
bounds $L_{\pm}$ and corresponding $H_0$-curves for flat
models.\footnote{In the case of open models $k=-1$ Eq. (6) allows
solutions if $Z<0$. These solutions are not discussed in present
paper.}

\begin{figure}[htb!] \centering{
\epsfig{file=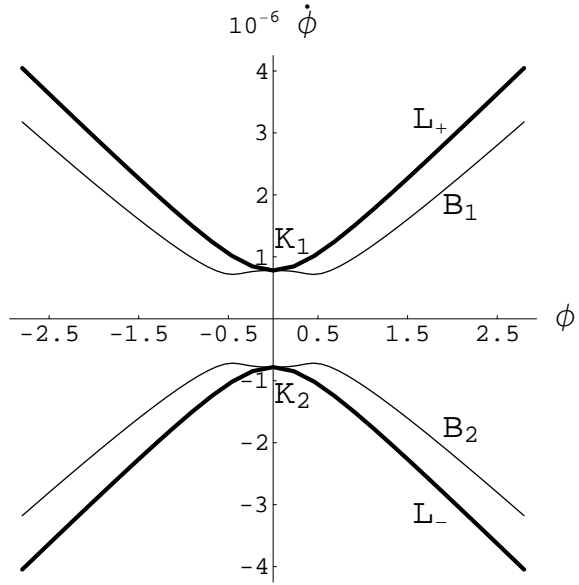,width=\linewidth} }
 \caption[]{$H_0$-curves of flat models at $\beta=-1{.}6\cdot
 10^{12}$}
\end{figure}

\begin{figure}[htb!]
\centering{ \epsfig{file=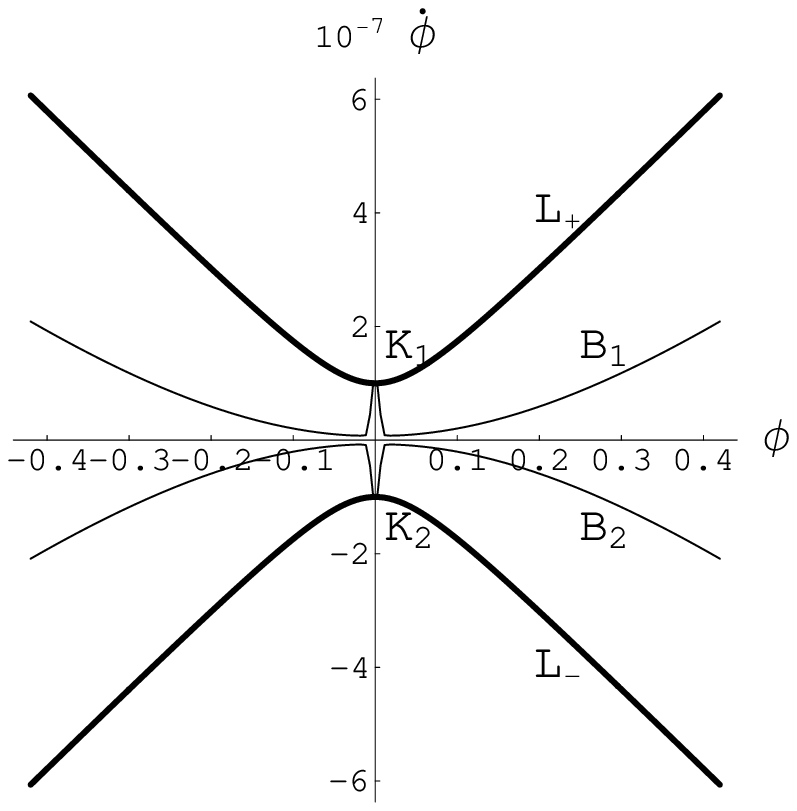,width=\linewidth} }
\caption[]{$H_0$-curves of flat models at $\beta=-1{.}0\cdot
 10^{14}$}
\end{figure}

\begin{figure}[htb!]
\centering{ \epsfig{file=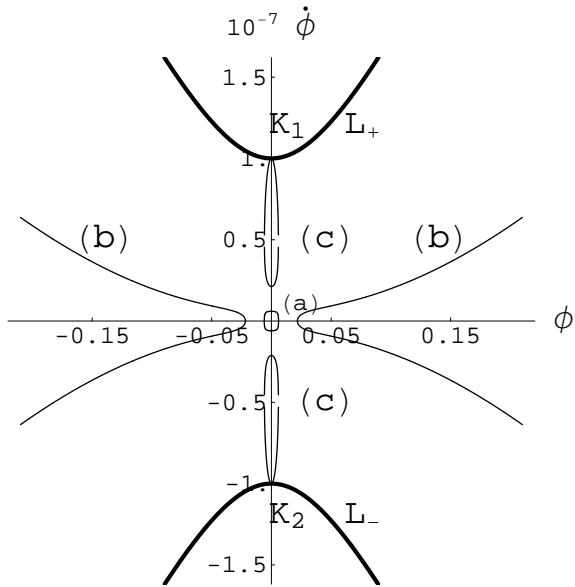,width=\linewidth} }
\caption[]{$H_0$-curves of closed models at $\beta=-1{.}0\cdot
 10^{14}$}
\end{figure}

For given scalar field potential $V(\phi)$ the behaviour of
$H_0$-curves depends on parameters of potential $V$, on parameter
$\beta$ and on the model type ($k=0,\,\pm1$). Below $H_0$ are
analyzed for scalar field potential $V=\frac{1}{2}m^2\phi^2$
($m=10^{-6}M_p$) in the case of $\rho_r=0$. For flat models
$H_0$-curves are given in Fig.~1 ($\beta=-1{.}6\cdot 10^{12}$).
The derivative $\dot{H}_0$ is positive in points of $H_0$-curves
if the value of $|\beta|<|\beta_{1}|\sim 1.8\cdot 10^{12}$. This
means that $H_0$-curves in considered case are bounce curves
discussed in \cite{l8}. By increasing of the value of $|\beta|$
the points $K_1$ and $K_2$ approach to the origin of coordinates
and the derivative $\dot{H}_0$ becomes negative in the
neighbourhood of points $K_1$ and $K_2$ (see Fig.~2). In points of
$H_0$-curves with $\dot{H}_0<0$ the evolution of system
corresponds to transition from expansion to compression. The
existence of regions on $H_0$-curves with $\dot{H}_0<0$ is
necessary condition for appearance of oscillating solutions for
SGS. The behaviour of $H_0$-curves for open models is similar to
that for flat models. The curves $B_1$ and $B_2$ are limiting
$H_0$-curves for open models under $R_0\to \infty$. By decreasing
of the value of $R_0$ corresponding $H_0$-curves of open models
approach to bounds $L_{\pm}$. All $H_0$-curves of open models
contain points $K_1$ and $K_2$, where $H_{+}=H_{-}$. Like to flat
models oscillating SGS-solutions of open type can appear by
sufficiently large values of $|\beta|$.

In the case of closed models by certain value of $\beta$ there are
three kinds of $H_{0}$-curves (Fig.~3). For very large values of
$R_0>R_{01}$ $H_0$-curves are closed curves with the center in
origin of coordinates (curve $(a)$ in Fig.~3), they correspond to
$H_0$-curves of GR. The value of $R_{01}$ depends on $|\beta|$,
for example $R_{01}\sim 7{.}14\cdot 10^7\,M_p^{-1}$ for
$|\beta|\sim 1{.}0\cdot 10^{14}\, M_p^{-4}$. In the case of
$R_0<R_{01}$ we have $H_0$-curves $(b)$ and $(c)$ presented in
Fig.~3. The Hubble parameter in points of curves $(b)$ is positive
(bounce-curves), and it is negative in points of curves $(c)$. So,
in points of $H_0$-curves ($\phi_0\neq 0$, $\dot{\phi}_0=0$) the
values of $R_0$ and $V_0$ are connected by usual relation of GR
$\displaystyle R_0^{-2}=\frac{8\pi}{3M_p^2}\,V_0$, and the
derivative $\displaystyle \dot{H}_0=\frac{8\pi}{3M_p^2}\,V_0-
\frac{3\beta {V'}_0^2}{1-4\beta V_0}>0$. In points of $H_0$-curves
on the axis $\dot\phi$ ($\phi_0=0$, $\dot{\phi}_0\neq 0$) we have
\begin{eqnarray}
&&R_0^{-2}=\frac{4\pi}{3M_p^2}\,\frac{\dot{\phi}_0^2}{Z_0}\left(
1-\frac{1}{2}\beta\dot{\phi}_0^2\right)\nonumber \\
&&\left(Z_0=1+\beta\dot{\phi}_0^2>0\right)\nonumber\\
\lefteqn{\text{and}}\nonumber\\
 & & \hspace{-2em}
\dot{H}_0=-\frac{\phi_0^2}{1-2\beta\dot{\phi}_0^2}\,
\left[\frac{8\pi}{3M_p^2}\left(1+\frac{1}{4}\beta\dot{\phi}_0^2\right)
-3\beta V''_0\right]<0.\nonumber
\end{eqnarray}
The presence on $H_0$-curves of regions with different signs of
$H_{+}$ ($H_{-}$) leads to appearance of oscillating solutions.
Note, that in points of $H_0$-curves lying on coordinates axes we
have  $H_{+}=H_{-}$.

The analysis shows that the behaviour of $H_0$ in the case of
potential $V_2=\frac{1}{4}\lambda \phi^4$
($\lambda=\mathrm{const}$) qualitatively is like to that of
considered above potential $V_1=\frac{1}{2}m^2\phi^2$.

\section{Numerical analysis of solutions for scalar fields  SGS}

By using various scalar field potentials applying in inflationary
cosmology (in particular, $V_1=\frac{1}{2}m^2 \phi^2$,
$V_2=\frac{1}{4}\lambda\phi^4$), numerical analysis of SGS
solutions of Eqs. (1), (2), (4) for flat, open and closed type was
carried out, solutions properties in dependence on value of
parameter $\beta$ and initial conditions were investigated. As
result the following conclusions were obtained:
\begin{enumerate}
\item SGS-solutions of flat and open type have non-stationary
character and are unstable. Particular SGS-solution of flat type
is  given in Fig.~4.\footnote{All numerical solutions in Fig.~4--6
are given in the case of potential  $V_1=\frac{1}{2}m^2 \phi^2$
($m=1{.}0\cdot 10^{-6}M_p$) by using the system of units with
$\hbar=c=M_p=1$.}
\item SGS-solutions of closed type can have quasi-stationary as
well as non-stationary regime in dependence on the value of
$\left|\beta\left(4V_0-\dot{\phi}_0^2\right)\right|$. The value of
$\phi_0^2$ has to be less than $4V_0$ at several orders to
conserve the stable character of discussed solutions. In
connection with this the analysis of SGS-solutions will be given
below in dependence on the value of $x=4|\beta V_0|$.
\item Quasi-stationary regime for SGS-solutions of closed type
takes place, if the value $x\sim 1$ and changes in certain
interval $x_{\mathrm{min}}\le x\le x_{\mathrm{max}}$ depending on
the value of $\beta$ and the potential form. So, in the case of
potential $V_1$ we have $1.5\le x\le 4.5$ at $|\beta|=1{.}0\cdot
10^{14}$. By increasing of $|\beta|$ the values of
$x_{\mathrm{min}}$ and $x_{\mathrm{max}}$ decrease. For example,
in the case of $V_1$ and $|\beta|=1{.}0 \cdot 10^{20}$ we have
$0{.}12\le x \le 2{.}0$. The frequency of $\phi$-oscillations and
also frequency and amplitude of $H$-oscillations do not depend
practically on the value of $\beta$ for given form of scalar field
potential in the case of quasi-stationary regime. The realization
of quasi-stationary regime means, that the sign of term
$3H\dot{\phi}$ in Eq. (4) changes during $\phi$-oscillations
period, during one half period this term plays the role of damping
force and during the second half period it plays the role of
accelerating  force. As result, the frequency of $H$-oscillations
is two times greater than the frequency of $\phi$-oscillations. In
accordance with (3) torsion (nonmetricity) function oscillates
with the same frequency as the Hubble
parameter\footnote{Quasi-stationary oscillating solutions for
closed models were discussed in \cite{l9} with purpose to build
non-singular cosmological model. From our point of view
satisfactory cosmological model can not be built by using
solutions for scalar fields SGS.}. The evolution of magnitudes
$\phi$, $H$, $R$ and the energy density of scalar field
$\rho_\phi=\frac{1}{2}\dot{\phi}^2+V(\phi)$ is presented in
Fig.~5. Note, that scalar field energy of closed SGS changes with
the time also.
\item By decreasing of the value of $x_{\mathrm{min}}$
($x<x_{\mathrm{min}}$) oscillating SGS-solutions of closed type
become unstable.
\item If $x\gg 1$ oscillating SGS-solutions of closed type are essentially
non-stationary for limited time intervals (Fig.~6). In the case of
large time scale oscillations have quasi-modulated character.
\item In accordance with the structure of GCFE the presence of
radiation has essential influence on dynamics of SGS in the case
of quasi-stationary regime $x\sim 1$ destroying this regime, and
such influence is essentially less in the case of non-stationary
regime $x\gg 1$.
\end{enumerate}
In conclusion note, that oscillating classical solutions for
scalar fields are unstable with respect to quantum processes of
transformation of scalar fields into elementary particles and
hence such solutions can have physical sense only for limited time
intervals.

\begin{figure}[htb!]
\begin{minipage}{0.47\textwidth}\centering{
\epsfig{file=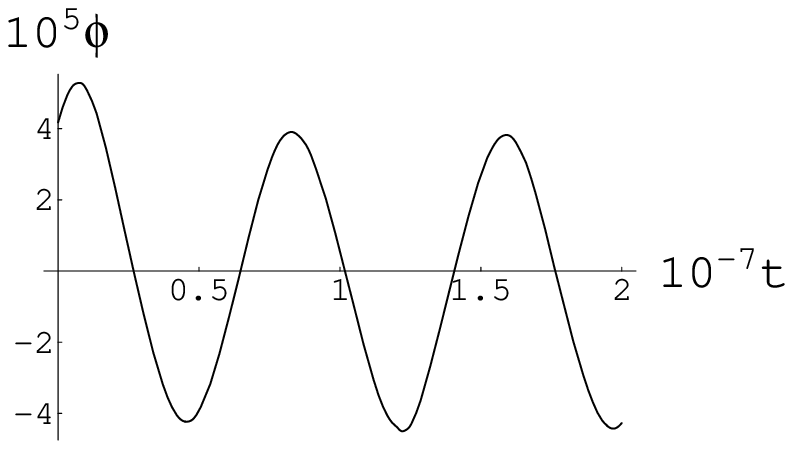,width=\linewidth} }
\end{minipage}\, \hfill\,
\begin{minipage}{0.47\textwidth}\centering{
\epsfig{file=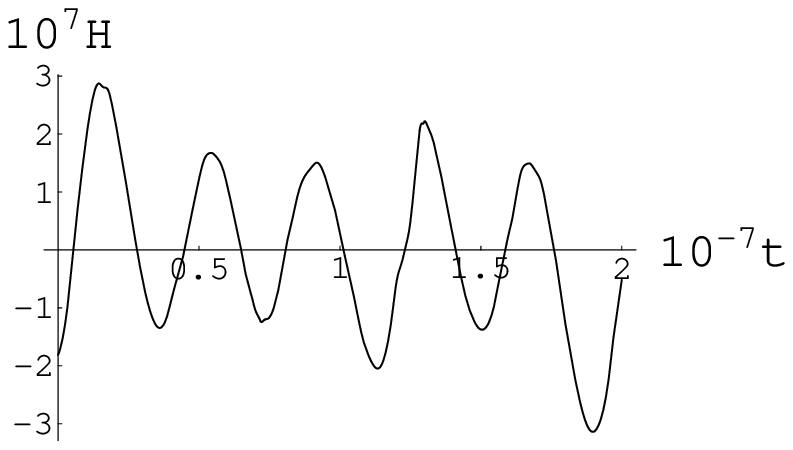,width=\linewidth}}
\end{minipage}\, \hfill\,
\caption[]{SGS-solution of flat type ($\beta=-1{.}0\cdot 10^{20}$,
$\phi_0=5{.}25\cdot 10^{-5}$ and $\dot{\phi}_0=7{.}09\cdot
10^{-12}$).}
\end{figure}
{\allowdisplaybreaks
\begin{figure}[htb!]
\begin{minipage}{0.47\textwidth}\centering{
\epsfig{file=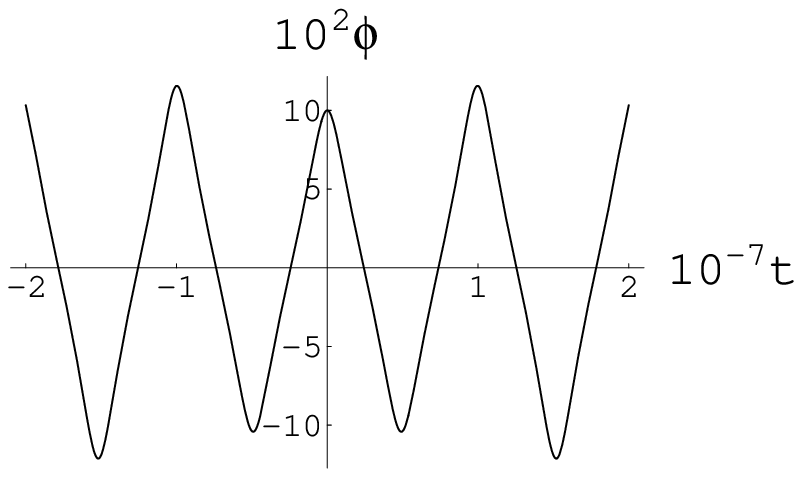,width=\linewidth} }
\end{minipage}\, \hfill\,
\begin{minipage}{0.47\textwidth}\centering{
\epsfig{file=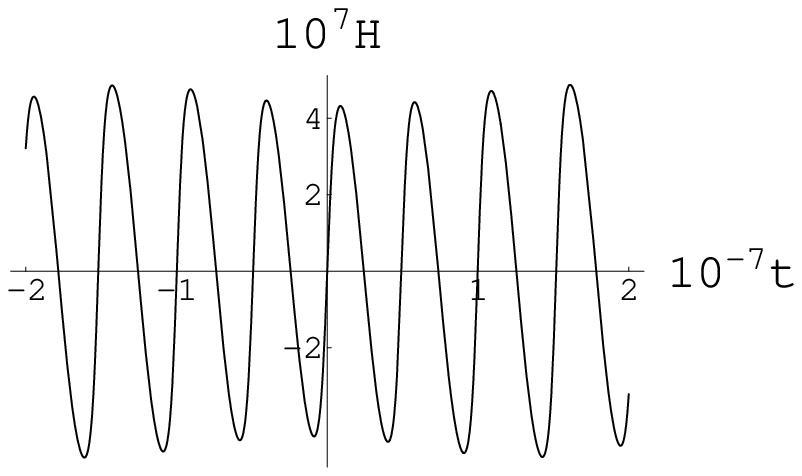,width=\linewidth}}
\end{minipage}\, \hfill\,\\
\begin{minipage}{0.47\textwidth}\centering{
\epsfig{file=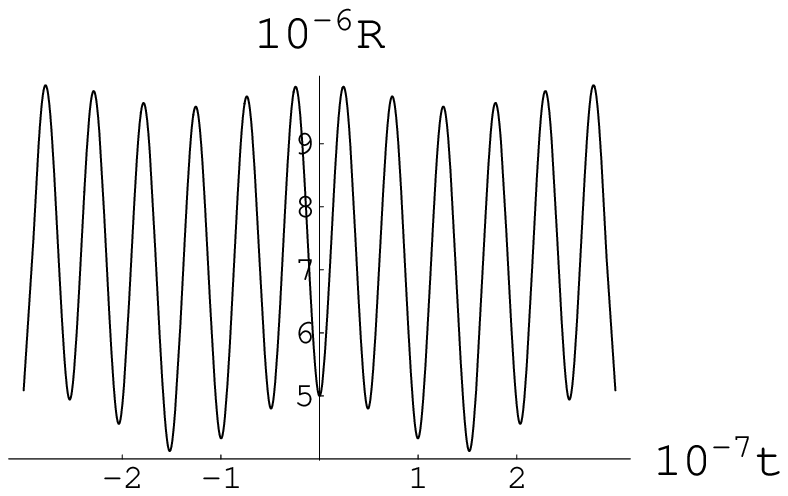,width=\linewidth} }
\end{minipage}\, \hfill\,
\begin{minipage}{0.47\textwidth}\centering{
\epsfig{file=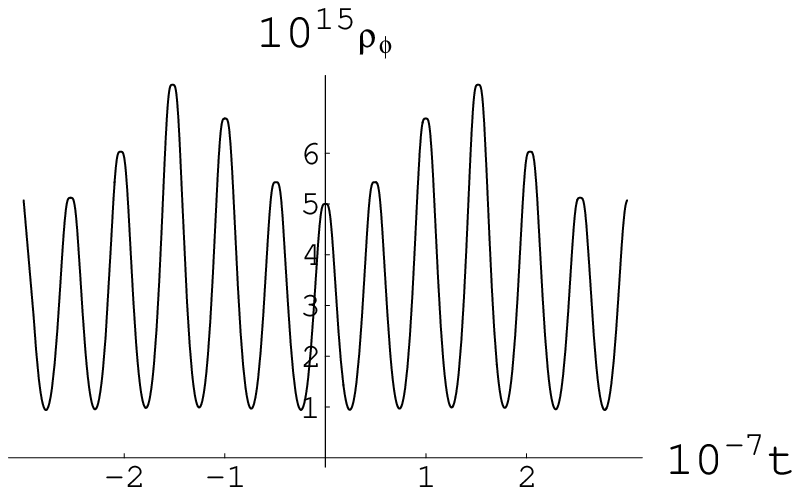,width=\linewidth}}
\end{minipage}\, \hfill\,
\caption[]{SGS-solution of closed type in quasi-stationary regime
($\beta=-1{.}0\cdot 10^{14}$, $\phi_0=0{.}1$ and
$\dot{\phi}_0=0$).}
\end{figure} }
\begin{figure}[htb!]
\begin{minipage}{0.47\textwidth}\centering{
\epsfig{file=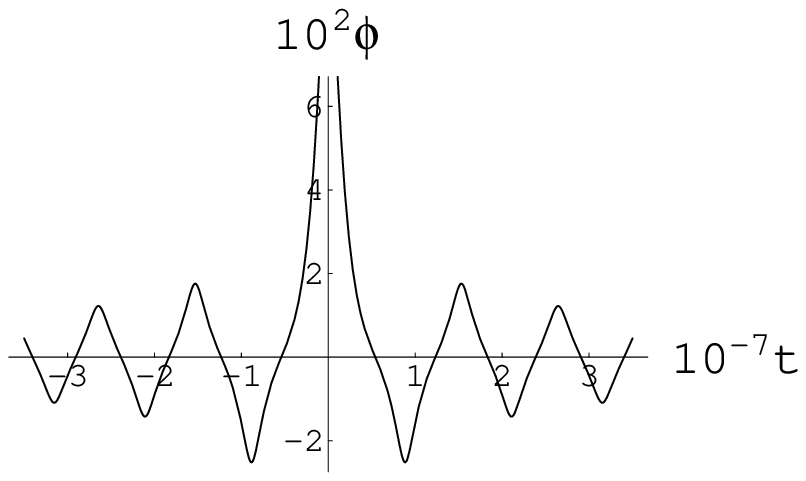,width=\linewidth} }
\end{minipage}\, \hfill\,
\begin{minipage}{0.47\textwidth}\centering{
\epsfig{file=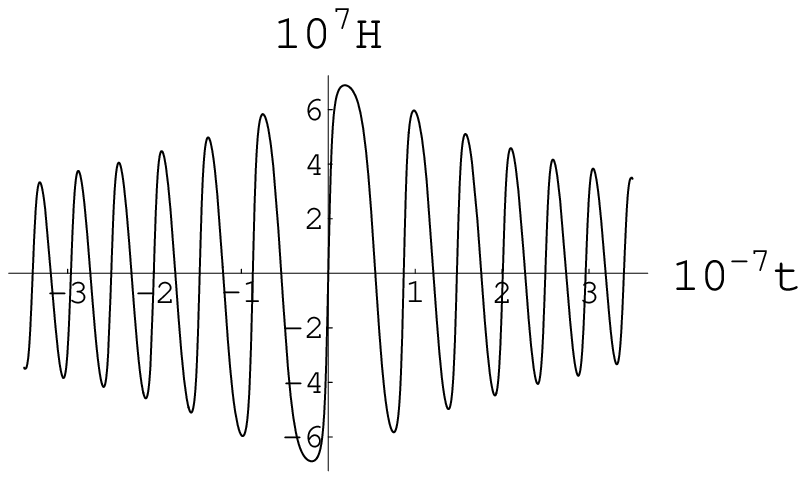,width=\linewidth}}
\end{minipage}\, \hfill\,
\caption[]{SGS-solution of closed type in non-stationary regime
($\beta=-1{.}0\cdot 10^{16}$, $\phi_0=0{.}1\cdot 10^{-5}$ and
$\dot{\phi}_0=0$.)}
\end{figure}

\end{document}